\documentclass[twocolumn,showpacs,preprintnumbers,amsmath,amssymb]{revtex4}
\usepackage{graphicx}
\usepackage{dcolumn}
\def\v#1{\mbox{\boldmath $#1$}}

\begin{document}
\title{Diffusiophoresis in a Near-critical Binary Fluid Mixture}


\author{Youhei Fujitani}
 \email{youhei@appi.keio.ac.jp}
\affiliation{School of Fundamental Science and Technology,
Keio University, 
Yokohama 223-8522, Japan}

\date{\today}

\begin{abstract}
We consider placing a rigid spherical particle into a binary fluid mixture
in the homogeneous phase near the demixing critical point.
The particle surface is assumed to 
have a short-range interaction with each mixture component and
to attract one component more than the other.  Owing to large osmotic susceptibility, 
the adsorption layer, where the preferred component is more concentrated, can be
{of significant thickness}.   {This causes a particle motion 
under an imposed composition gradient.}  Thus, diffusiophoresis emerges from a mechanism
which has not been {considered so far}. 
We calculate how the mobility depends on the temperature and particle size.
\end{abstract}

\maketitle
Diffusiophoresis \cite{derja,andersJFM,SepPurMeth,prieve,anders,staff}
{has} recently attracted much attention {mainly because of} their applications in lab-on-a-chip processes 
\cite{shin,marb,keh,abecas,volpe,veleg,osmflow,kar,shinPNAS,frenkel,vinze}.
They  are conventionally discussed in terms of a solution in contact with a solid.
A solution in the interfacial layer, generated immediately near a solid surface, can have distinct properties
due to some interaction potential between a solute and a surface.
When a gradient of solute concentration is imposed in the direction tangential to the surface,
some force exerted on only the solution in the interfacial layer
can yield the slip velocity between the solid and the bulk of the solution.
Diffusiophoresis occurs when a solution has no convection far from a mobile
surface, whereas diffusioosmosis occurs when a surface is fixed.  
If the solution is an electrolyte, the interfacial layer is the electric double layer.
Otherwise, the layer can be generated by the van der Waals interaction or some dipole-dipole interaction.
\par

Below, a binary fluid mixture, containing no ions, in the homogeneous phase close to the demixing critical point
is considered and is simply referred to as a mixture.   We can apply hydrodynamics 
for flow with a typical length sufficiently large in comparison to the  correlation length
of {composition fluctuations, which
are} significant only on smaller length scales \cite{okafujiko, furu}.  
Suppose that a rigid spherical particle is placed into a mixture with the critical composition. 
A short-range interaction is assumed between each mixture component and
the particle surface.  The surface usually 
attracts one component  more than the other  
\cite{Cahn, binder, holyst, beyslieb, beysest,diehl97}, although this is not always the case
\cite{beys2019, fujitani2021}.  
The adsorption layer, where the preferred component is more concentrated,
can {be of significant thickness} because of large osmotic susceptibility.
Assuming a mixture to be quiescent far from the particle, 
we previously calculated how the preferential adsorption (PA) affects
the drag coefficient $\gamma_{\rm d}$, which is
defined as the negative of the ratio of the drag force to a given particle velocity in the linear regime \cite{yabufuji}.  
\par

We write $T$ for the temperature, $T_{\rm cr}$ for the critical temperature, and  $\tau$
for {the reduced temperature} $|T-T_{\rm cr}|/T_{\rm cr}$.
In the renormalized local functional theory \cite{fisher-auyang,OkamotoOnuki}, 
the free-energy functional (FEF) is coarse-grained 
up to the local correlation length, $\xi$.
After coarse-grained, the composition's equilibrium profile 
minimizes the FEF 
because many profiles that differ only on smaller length scales are unified into much fewer profiles \cite{OkamotoOnuki, Yabu-On}.  
In our previous work \cite{yabufuji}, the hydrodynamics formulated from
the coarse-grained FEF \cite{yabuokaon,undul} is used, and  the {critical} composition is {assumed}  
far from the particle.  {There,} 
$\xi$ becomes equal to $\xi_0 \tau^{-\nu}$, which is denoted by $\xi_\infty$.  
Here, $\xi_0$ is a non-universal length of molecular size and $\nu\ (\approx 0.630)$
is a critical exponent \cite{peli}. {We assume $\xi\gg \xi_0$}.  The PA significantly affects
 $\gamma_{\rm d}$ if the particle radius, $r_0$, is not much larger than $\xi_\infty$, which 
can indicate the thickness of the adsorption layer \cite{yabufuji}.  
{Even then, }$\xi$ can be sufficiently small anywhere as compared with a typical length of the flow
{because of $\xi<\xi_\infty$ in the adsorption layer}.  Experimentally, $\xi_\infty$ can reach $100\ $nm. 
\par

\begin{figure}
\includegraphics[width=6cm]{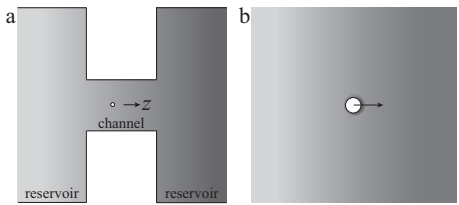}
\caption{Schematic of the situation supposed in our formulation.
The grayscale indicates the composition;
a component is more concentrated in a darker region.  
(a)  A mixture is filled in a container, where a channel connects two reservoirs.  
A particle, represented by a circle, passes through the channel,
which extends along the $z$ axis.
 (b) A magnified view of a region including the particle.  
A darker region immediately near the particle
represents the adsorption layer. The particle motion is represented by an arrow.
}
\label{fig:container}
\end{figure}

In this Letter, we use the hydrodynamics mentioned above to discuss diffusiophoresis caused by
the PA onto the particle surface.
As shown in Fig.~\ref{fig:container}, a neutrally buoyant particle lies in
a mixture filled in a channel, which has
dimensions much larger than $r_0$ and extends along the $z$ axis.
We impose {a} composition difference between the reservoirs to create a 
weak composition gradient in the channel.  The 
particle undergoes a stationary and translational motion free from the drag force. 
If the PA were also assumed onto the channel wall, the composition gradient would convect
the mixture via diffusioosmosis \cite{wolynes,pipe}.
In Fig.~\ref{fig:container}(b), we take a mixture region including the particle.
Far from the particle in this region, which has dimensions much larger than $\xi_\infty$,
the composition gradient becomes a constant vector
and convection vanishes. 
 In the following paragraphs, we will formulate this situation and
calculate mobility, which represents the ratio of the particle velocity to the constant vector.
An anisotropic stress in the adsorption layer causes diffusiopheresis and arises from 
the free energy in the bulk of a mixture, which 
does not involve any potential due to the surface
because the length range each component interacts with the surface
is invisible in our coarse-grained description. 
This mechanism is not considered in the conventional theory.
In our problem,
larger osmotic susceptibility should tend to increase the magnitude of mobility by thickening the adsorption layer, 
and have the opposite tendency by weakening the chemical-potential gradient under
a given composition gradient.  As a result, it is expected that the change in mobility with {the reduced temperature} $\tau$ can be
nonmonotonic.   The change with {the particle radius} $r_0$ is expected to be 
explicit unless $\xi_\infty /r_0$ is much smaller than unity.  
Our results are later shown to be consistent with {the} expectations.  {These properties}
may be utilized in manipulating colloidal particles.
\par

{Naming} the mixture components a and b, we write $\rho_n$ for
the mass density of the component $n\ (=$ a or b$)$, and $\mu_n$ for
the chemical potential conjugate to $\rho_n$.
We write $\rho$ for $\rho_{\rm a}+\rho_{\rm b}$,  
$\varphi$ for $\rho_{\rm a}-\rho_{\rm b}$,
and $\varphi_{\rm c}$ for the value of $\varphi$ at the critical composition.
Hereafter, a quantity at the critical composition is generally indicated by the subscript $_{\rm c}$.
The order parameter is given by $\psi\equiv \varphi-\varphi_{\rm c}$.
The chemical potential conjugate to $\varphi$ is given by $(\mu_{\rm a}-\mu_{\rm b})/2$,
and its deviation from the value at the critical point is denoted by $\mu$,
{which is simply called the chemical potential here.}
We write $c_n$ for the mass fraction of the component $n$;
$c_n=\rho_n/\rho$ and $\varphi=\rho(2c_{\rm a}-1)$ hold. 
We use the conventional symbols for the critical exponents,
$\beta,\ \gamma$, and $\eta$, {and have} $\eta{\approx 0.0364}$ \citep{peli}.
The (hyper)scaling relations give $2\beta+\gamma=3\nu$ and
$\gamma=\nu(2-\eta)$.  

\par
The coarse-grained FEF is separated into the $\rho$-dependent part and $\psi$-dependent part.
The latter part in the bulk of a mixture, denoted by ${\cal F}_{\rm b}[\psi]$, is
given by the volume integral of
\begin{equation}
f_{\rm b}(\psi)+ \frac{1}{2} K(\psi) \left|\nabla \psi\right|^2-\mu \varphi \label{eqn:fbplus}
\end{equation}
over the mixture. Derived in Ref.~\cite{OkamotoOnuki},
$f_{\rm b}{(\ge 0)}$ and $K{(>0)}$ {are even functions of $\psi$}, as
shown in the supplementary material.
The corresponding integrand for the $\rho$-dependent part is independent of the derivatives of $\rho$.  
The difference in the components' interactions with the surface contributes
to the FEF and is described by the
area integration of a function of $\varphi$ over the particle surface.  
Assuming it to be a linear function, we write $h$ for  
the negative of the coefficient of $\varphi$; $h$ is called the surface field.   
The sum of the area integration
and ${\cal F}_{\rm b}[\psi]$ gives the $\psi$-dependent part, denoted by ${\cal F}[\psi]$.
At equilibrium, $\mu$ is homogeneous.
In general,  $\mu$ can depend on the position ${\v r}$, $\psi$ minimizing ${\cal F}[\psi]$
gives the coarse-grained profile, and  the minimum of
${\cal F}[\psi]$ gives a part of the grand potential.
\par
 
The reversible part of the pressure tensor originating from 
${\cal F}_{\rm b}$ is denoted by ${\mathsf \Pi}$.
We can define $p$ so that $p {\mathsf 1}$ is the one from
the $\rho$-dependent part.  Here, ${\mathsf 1}$ denotes
the second-order isotropic tensor whose components {with respect to rectangular coordinates} 
are represented by Kronecker's delta, and
$p$ is the scalar pressure in the absence of ${\cal F}$. 
The reversible part of the pressure tensor is thus given by 
$p{\mathsf 1}+{\mathsf \Pi}$ in the bulk of a mixture.  
As in the model H \cite{Hohen-Halp},
${\mathsf \Pi}$ is obtained by subtracting the product of ${\mathsf 1}$ and
Eq.~(\ref{eqn:fbplus}) from $K\nabla \psi\nabla\psi$ \cite{onukibook,onukiPRE}.  
If $\psi$ is homogeneous, the scalar pressure, denoted by $P$,  
equals $p-f_{\rm b}(\psi)+\mu\varphi$.  

\par
In the reference state, {where} a particle {is} at rest in an equilibrium mixture
with the critical composition far from the particle, {the chemical potential} $\mu$ vanishes and 
{the total mass density} $\rho$ is homogeneous.  
A difference in {the composition} $c_{\rm a}$ between the reservoirs is imposed as a perturbation on this state. 
Assuming the difference to {be proportional to}
a {dimensionless} smallness parameter $\varepsilon$, we
{consider} the velocity of a force-free particle up to the order of $\varepsilon$. 
In a frame fixed to the container,
we define $U$ so that the particle velocity equals $\varepsilon U{\v e}_z$ up to the order of $\varepsilon$,
where  ${\v e}_z$ is the unit vector along the $z$ axis. 
{With} ${\v v}$ {denoting} the velocity field of a mixture, the no-slip boundary condition
{gives} ${\v v}=\varepsilon U{\v e}_z$ at the particle surface.  
Far from the particle in Fig.~\ref{fig:container}(b), because {$\nabla P={\v 0}$ is assumed there} and
no PA is assumed onto the channel wall, 
interdiffusion  {occurs between} the components under ${\v v}={\v 0}$.  
The composition gradient there is proportional to $\nabla \mu$ multiplied by the osmotic susceptibility.
\par

Up to the order of $\varepsilon$, {because} $\rho$ is stationary
in the frame co-moving with the particle center, {w}e can assume 
the incompressibility condition $\nabla\cdot{\v v}=0$.  
In a frame fixed to the container, the time derivative of $\varphi$ is equal to the negative of the divergence of 
the sum of the convective flux, $\varphi {\v v}$, and the diffusive flux.
The latter is given by $-L(\psi)\nabla\mu$, where a function $L(\psi)$ represents the transport coefficient for the interdiffusion.  
We thus have
$\left({\v v}-\varepsilon U{\v e}_z\right) \cdot \nabla \psi= \nabla \cdot \left[ L(\psi)\nabla \mu\right]$
with the aid of the stationarity in the co-moving frame up to the order of $\varepsilon$.
We consider a particular instant, when the particle center passes
the origin of the polar coordinate system $(r,\theta,\phi)$ fixed to the container.
The $z$ axis is taken as the polar axis.
A mixture does not diffuse into the particle, which yields $\partial_r \mu=0$ at $r=r_0$.
Here, $\partial_r$ indicates the partial derivative with respect to $r$.

\par
Correlated clusters of the order parameter {are convected} on 
length scales smaller than $\xi$, {which} enhances $L$ \cite{kawasaki, seng}.  
 The quotient of $L(0)$ divided by the osmotic susceptibility gives the diffusion coefficient
{for} the relaxation of {equilibrium order-parameter fluctuations at}  
the critical composition.
According to the mode-coupling theory \cite{kawasaki,onukibook},
the quotient {equals} the self-diffusion coefficient obtained by applying the Stokes law \cite{stokes} and Sutherland-Einstein relation
\cite{suther, eins} for a rigid sphere having a radius equal to $\xi$.
We  simply extend this result to a homogeneous off-critical composition,
and use the extended result, $L(\psi) = k_{\rm B}T / [{6\pi\eta_{\rm s}\xi f_{\rm b}''(\psi)}]$, 
in the dynamics {\cite{yabufuji}.  
Here,} $\xi$ and $\psi$ {imply} their respective local values {and} $k_{\rm B}$ 
{denotes} the Boltzmann constant.
The shear viscosity, denoted by $\eta_{\rm s}$, is assumed to be constant, with its weak singularity neglected.
The (double) prime represents the (second-order) derivative with respect to the variable. 
The osmotic susceptibility, with $T$ and $\rho$ kept constant, is given by $1/f_{\rm b}''(\psi)$,
which is proportional to $\tau^{-\gamma}$ for $\psi=0$ in the critical regime.
{We use} $\nabla\cdot {\mathsf \Pi}=\varphi \nabla\mu$ {and}
the Stokes approximation to {obtain}
$0=-\nabla p-\varphi\nabla\mu +\eta_{\rm s} \Delta {\v v}$ {from the law of} momentum conservation.
\par

A superscript $^{(0)}$ is added to a field in the reference state,
{where the order parameter can be written as} 
$\psi^{(0)}(r)$ because of the symmetry.
We define $p_{+}$ as $p+\varphi_{\rm c}\mu$.
Up to the order of $\varepsilon$, we expand the fields as
{$\psi({\v r})=\psi^{(0)}(r)+\varepsilon \psi^{(1)}({\v r})$,} 
${p}_+({\v r})={p}_+^{(0)}+\varepsilon {p}_+^{(1)}({\v r})$, $\mu({\v r})=\varepsilon \mu^{(1)}({\v r})$, 
and ${\v v}({\v r})=\varepsilon {\v v}^{(1)}({\v r})$.
A field with the
superscript $^{(1)}$ is defined so that it becomes
at the order of $\varepsilon$ after being multiplied by
$\varepsilon$.  Far from the particle, $p_+^{(1)}$ becomes constant
because there $\varepsilon p_+^{(1)}$ equals $P$ at the order of $\varepsilon$.
Up to the order of $\varepsilon$,  the interdiffusion is described by
\begin{equation}
\left({\v v}^{(1)}-U{\v e}_z\right) \cdot \nabla \psi^{(0)}= \nabla \cdot \left[ L(\psi^{(0)})\nabla \mu^{(1)}\right]
\ ,\label{eqn:diffusion1}\end{equation}
{which equals ${\psi^{(0)}}'L'\partial_r\mu^{(1)}+L\Delta \mu^{(1)}$,} 
and the momentum conservation {is described} by
\begin{equation}
0=-\nabla p_+^{(1)}-\psi^{(0)}\nabla\mu^{(1)}+\eta_{\rm s}\Delta {\v v}^{(1)}
\ .\label{eqn:sto1}\end{equation}
We define $\tau_*$ as $(\xi_{0}/{r_0})^{1/\nu}$ to 
use ${\hat \tau}\equiv \tau/\tau_*$, and define $L_*$ as $L(0)$ at $\tau=\tau_*$ to have
$L(0)
 = {\hat \tau}^{\nu\eta-\nu} L_*$.
Because of the particle-motion symmetry, the angular dependence of $\mu^{(1)}$ and that of 
$v_r^{(1)}$ are only via $\cos{\theta}$, 
that of $v_\theta^{(1)}({\v r})$ only via $\sin{\theta}$, and $v_\phi^{(1)}$ vanishes \cite{fpd}.  
A spatially constant term in $\mu$, if any, does not
affect our calculation and can be neglected.  
We {nondimensionalize $\mu^{(1)}$ and $v_r^{(1)}$ to define}
functions ${\cal Q}$ and ${\cal R}$ so that
\begin{equation}
{\cal Q}({\hat r})\cos{\theta} =  \frac{\sqrt{L_*}\mu^{(1)}({\v r})}{U \sqrt{5\eta_{\rm s}}} 
\quad{\rm and}\quad {\cal R}({\hat r})\cos{\theta}=\frac{v_r^{(1)}({\v r})}{U} 
\label{eqn:defQR}
\end{equation}
hold, where ${\hat r}$ denotes the dimensionless radial distance $r/r_0$. 
Writing ${\cal D}_{m}$ {for} 
${\hat r}\partial_{\hat r}+m$ and {defining}
$\Phi({\hat r})$ as
$-{r^2} {\psi^{(0)}}'(r_0{\hat r}) / [ 3 \sqrt{5\eta_{\rm s} L_*} ]$,
we rewrite Eq.~(\ref{eqn:diffusion1}) as 
\begin{equation}
{\cal D}_{-1}{\cal D}_{2}
{\cal Q}({\hat r})
=-3\Phi({\hat r})\left[{\cal A}({\hat r})\left({\cal R}({\hat r})-1\right) -{\cal B}({\hat r}){\cal Q}'({\hat r})
 \right]\ ,
\label{eqn:calQ}\end{equation}
{whose left-hand side comes from 
$\Delta \mu^{(1)}$.
  In the supplementary material,
definitions of ${\cal A}$ and ${\cal B}$ and some equilibrium order-parameter profiles
are presented.}  From {the incompressibility condition and the $r$ and $\theta$ components of} Eq.~(\ref{eqn:sto1}),    
we eliminate $p_+^{(1)}$ 
{and $v_\theta^{(1)}$} to obtain
\begin{equation}
{\cal D}_{1}{\cal D}_{-2}{\cal D}_{3}{\cal D}_{0} 
{\cal R}({\hat r})
=-30 \Phi({\hat r}) {\cal Q}({\hat r})\ .
\label{eqn:calR}\end{equation}  
These two equations are derived in Ref.~\cite{yabufuji}.
The boundary conditions for ${\cal Q}$  and ${\cal R}$, respectively derived from those for $\mu$ and ${\v v}$, 
are the same as given in Ref.~\cite{yabufuji} except that ${\cal Q}({\hat r})$ is proportional to ${\hat r}$
far from the particle.  We write ${g}$ for the constant of proportionality; Ref.~\cite{yabufuji} supposes ${g}=0$.  
{Far from the particle, 
$\nabla \mu$, and thus $\nabla c_{\rm a}$, are proportional to $\varepsilon {\v e}_z$ up to the order of $\varepsilon$.} 
%
\par

{Applying} the boundary conditions, we can transform Eq.~(\ref{eqn:calQ}) so that 
${\cal Q}$ equals the sum of the solution in the absence of the right-hand side (RHS) and 
the convolution of a kernel and the RHS.
We can transform Eq.~(\ref{eqn:calR}) similarly by using another kernel.
The kernels
are respectively denoted by $\Gamma_{\rm Q}$ and $\Gamma_{\rm R}$, which
are {obtained in Ref.~\cite{okafujiko}.}
As in {Ref.~\cite{yabufuji}}, we assume that the RHSs 
are multiplied by an artificial parameter $\kappa$.  Accordingly, {rewriting}
${\cal Q}$ as
${\tilde {\cal Q}}_\kappa$ and  ${\cal R}$ as ${\tilde {\cal R}}_\kappa$,
{we have 
simultaneous integral equations for
${\tilde {\cal Q}}_\kappa$  and ${\tilde {\cal R}}_\kappa$. They are shown in the supplementary material,
together with the definitions of the kernels.}
We find ${\tilde {\cal Q}}_0({\hat r})$ equal to
${g}({\hat r}+({\hat r}^{-2}/2))$, which is below referred to as $q_0({\hat r})$.
Eliminating ${\tilde {\cal R}}_\kappa$ from the integral equations,
and expanding ${\tilde {\cal Q}}_\kappa$ with respect to $\kappa$ as
${\tilde{\cal Q}}_\kappa ({\hat r}) = \sum_{n=0}^{\infty} {q}_{n}({\hat r})\kappa^{n}$, 
we obtain a recurrence relation for the expansion coefficients, $q_0, q_1, q_2, \ldots$, as
\begin{eqnarray}
&&{q}_{n}({\hat r})=\int_1^\infty d{\hat s}\ \Gamma_{\rm Q}\left({\hat r}, {\hat s}\right)
\Phi\left({\hat s}\right)\Big[ -{\cal B}\left({\hat s}\right)q_{n-1}'\left({\hat s}\right)\nonumber\\
&&\quad +{\cal A}({\hat s}) \int_1^\infty d{\hat u}\ 
\Gamma_{\rm R} \left({\hat s}, {\hat u}\right)
\Phi\left({\hat u}\right)q_{n-2}\left({\hat u}\right)\Big]
\label{eqn:qn}\end{eqnarray}
for $n=1,2,3, \ldots$.  Here, for $n=1$, 
the integral with respect to ${\hat u}$ is replaced by $\alpha_0({\hat s})\equiv   
 (3{\hat s}^{-1}/2)-({\hat s}^{-3}/2)-1$.
The original solutions of ${\cal Q}$ and ${\cal R}$ are respectively given by ${\tilde {\cal Q}}_1$ and ${\tilde {\cal R}}_1$.  
\par

The drag coefficient for a value of $g$ 
in general deviates from the Stokes law, $6\pi\eta_{\rm s}r_0$, 
owing to the PA and the imposed gradient. For the drag coefficient
considered here, {Eq}.~(3.28) of Ref.~\cite{yabufuji} 
{remains} valid. {Its derivation is outlined
in the supplementary material.} 
 Thus, writing $\Delta{\hat \gamma}_{\rm d}$ for
the ratio of the deviation to the Stokes law, we find  $\Delta{\hat \gamma}_{\rm d}$ to be given by the integral of
$10\alpha_0({\hat r})\Phi({\hat r}) {\cal Q}({\hat r})/3$ from ${\hat r}=1$ to $\infty$. 
Obtaining $q_n$ from Eq.~(\ref{eqn:qn}) and replacing ${\cal Q}$ by
$q_n$ in the integral above, we find that the sum of the integrals over $n=0,1,2\ldots$  
gives $\Delta{\hat \gamma}_{\rm d}$.  Unless otherwise stated, we truncate this series 
suitably to calculate $\Delta{\hat \gamma}_{\rm d}$ numerically, 
with the aid of the software Mathematica (Wolfram Research).
From the governing equations and the boundary conditions for ${\cal Q}$ and ${\cal R}$,
we find $\Delta{\hat \gamma}_{\rm d}$ to be a linear function of ${g}$.
The  linear function, which is obtained by calculating $\Delta{\hat \gamma}_{\rm d}$ for ${g}=0$ and $1$,
determines the value of ${g}$ making the drag force vanish, {{\it i.e.\/}, yielding $\Delta{\hat \gamma}_{\rm d}=-1$.}  
This value is denoted by $g_0$.  {We note that $g$ is
a dimensionless value of $\partial_z \mu^{(1)}$ far from the particle with the velocity $\varepsilon U{\v e}_z$.}
The sign of $\psi^{(0)}$ is changed by changing that of {the surface field} $h$, 
{and thus} we find $g_0$ to be an odd function of $h$.  
{In the absence of PA $(h=0)$, the second term on the RHS of Eq.~(\ref{eqn:sto1}) 
vanishes, $\Delta{\hat \gamma}_{\rm d}$ vanishes irrespective of $g$, $g_0$ cannot be obtained,
and diffusiophoresis does not appear.} 
\par

\begin{figure}
\includegraphics[width=7.5cm]{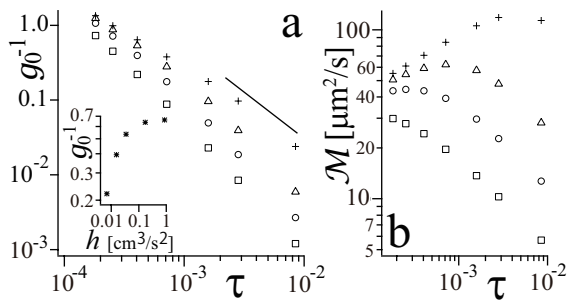}
\caption{{By setting the particle radius} $r_0$ {equal to} $0.1\ \mu$m. 
{we plot a dimensionless coefficient} $1/g_0$ in (a) and {the mobility} ${\cal M}$ 
in (b) against {the reduced temperature} $\tau$.
{Linked via Eq.~(\ref{eqn:mob}), they
are proportional to the particle speeds under given gradients of the chemical potential $\mu$ and 
the composition $c_{\rm a}$
far from the particle, respectively. The results shown by t}he symbols $+$, $\triangle$, $\circ$, and $\square$ 
{are respectively obtained by setting the surface field} $h$, {defined below Eq.~(\ref{eqn:fbplus}), equal to}
$1.73\times 10^{-1},\ 3.46\times 10^{-2},\ 1.55\times 10^{-2}$, and $6.91\times  10^{-3}\ $cm$^3/$s$^2$.
The reference line in (a) has the slope of $-\gamma$.
In the inset, $1/g_0$ is plotted against $h$ for $\tau=4.05\times 10^{-4}$ and $r_0=0.1\ \mu$m.
For its result on the extreme right, 
the series {for the deviation ratio of the drag coefficient} $\Delta{\hat \gamma}_{\rm d}$ numerically diverges and 
the Borel summation is applied instead of the suitable truncation.}
\label{fig:g0inv}
\end{figure}

{We first assume the chemical-potential gradient} $\nabla\mu
{\ \left( \propto \varepsilon {\v e}_z\right)}$ {to be} given
{f}ar from the particle.  The velocity of a force-free particle 
equals the product of the given gradient, $r_0\sqrt{L_*/5\eta_{\rm s}}$, and $1/g_0$,
because of the first entry of Eq.~(\ref{eqn:defQR}).
Below, we use the parameter values by supposing
a mixture of 2,6-lutidine and water (LW), which has
$T_{\rm cr}=307\ $K, $c_{\rm ac}=0.290$, and $\xi_0=0.198\ $nm \cite{mirz}.
In Fig.~\ref{fig:g0inv}(a),
we find for $h>0$ that $1/g_0$, being positive, increases as {the reduced temperature} $\tau$ decreases and 
as {the surface field} $h$ increases.
This increase of $1/g_0$ is expected because then the PA is stronger.
In the inset of Fig.~\ref{fig:g0inv}(a), the increase of $1/g_0$ with $h$ becomes more gradual as $h$ increases
for the value of $\tau$. 
In the supplementary material, 
{we show some flow profiles,} 
{which are consistent with} our coarse-grained description.  
\par

\begin{figure}[h]
\includegraphics[width=5cm]{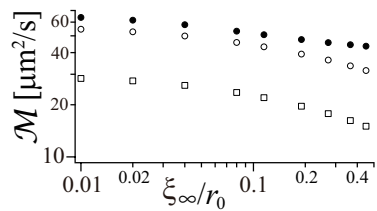}
\caption{{We plot the mobility} ${\cal M}$ against 
{a dimensionless correlation length far from the particle in the reference state}
$\xi_\infty/r_0$, {which indicates
a relative thickness of the adsorption layer.  As mentioned in the
supplementary material, $L_*$ in Eq.~(\ref{eqn:mob}) depends on 
the particle radius $r_0$.}
Solid and open circles
represent results for $\tau=1.82\times 10^{-4}$ and $7.14\times 10^{-4}$, respectively, under $h=
1.55\times 10^{-2}\ $cm$^3/$s$^2$.   For these values of $\tau$, $\xi_\infty$ equals $45.0$ and
$19.0\ $nm, respectively.  Squares represent results for $\tau=7.14\times 10^{-4}$ and 
$h=6.91\times  10^{-3}\ $cm$^3/$s$^2$.  }
\label{fig:mobxi}
\end{figure}

To consider the diffusiophoretic mobility, we {then} assume {the composition gradient}
$\nabla c_{\rm a}\ \left( \propto \varepsilon {\v e}_z\right)$ {to be} given far from the particle.
 Writing ${c_{\rm a}'}_\infty$ for the $z$ component of the given gradient, 
we define the mobility ${\cal M}$ so that
$\varepsilon U = {\cal M} {c_{\rm a}'}_\infty /c_{\rm ac}$ holds, and have
\begin{equation}
{\cal M}=\frac{r_0c_{\rm ac}\sqrt{L_*}}{g_0 \sqrt{5\eta_{\rm s}}} \times \left.\frac{\partial\mu}{\partial c_{\rm a}}\right)_{TP} 
\ .\label{eqn:mob}\end{equation}
Here, the partial derivative, with $(T, P)$ 
fixed, is evaluated far from the particle in the reference state.
It is positive because of the thermodynamic stability. Thus,
the particle moves toward the
reservoir where the component preferred by the particle surface is more concentrated,
as shown in Fig.~\ref{fig:container}(b).  The conventional diffusiophoresis has a similar property \cite{anders,staff}.
{A}s shown in the supplementary material, the partial derivative 
approximately equals $2\rho_{\rm c}f_{\rm b}''(0) \ (\propto\tau^\gamma)$. 
In Fig.~\ref{fig:g0inv}(b), the slope of ${\cal M}$ is negative 
at $\tau\approx 10^{-2}$ for each value of $h$.
Except for the smallest value of $h$, 
the slope is positive for small $\tau$ and
the value of $\tau$ at the peak of ${\cal M}$ increases with $h$. 
The slope becomes negative when the decrease of $1/g_0$ with increasing $\tau$,
{shown in} Fig.~\ref{fig:g0inv}(a), overcomes the increase of $\tau^\gamma$.
In Fig.~\ref{fig:mobxi}, the mobility ${\cal M}$ increases with {the particle radius} $r_0$, as in the conventional theory \cite{anders, mood}.
The increase is {more} noticeable when $r_0$ is smaller than approximately $0.2\ \mu$m at the larger value of $\tau$.  
\par

Suppose that the difference in $c_{\rm a}$ between the reservoirs is $c_{\rm ac}/10$
and that the channel length is $10\ \mu$m.  
The composition then remains close to the critical one over the channel.  
If we adopt ${\cal M}=50\ \mu$m$^2/$s from Fig.~\ref{fig:g0inv}(b), the particle speed is $0.5\ \mu$m$/$s
for $r_0=0.1\ \mu$m.  
{For Eq.~(\ref{eqn:sto1}) to be meaningful under $h\ne 0$,
$\psi^{(0)}$ is required to be much larger than $\varepsilon \psi^{(1)}$ somewhere. 
Considering that the perturbation involves the velocity field,
this requirement can be simplified in terms of a typical energy per unit area as $\eta_{\rm s} \left|\varepsilon U \right|\ll \left|h\psi^{(0)}(r_0)\right|$. 
This is satisfied} in the example presented here, as illustrated in the supplementary material. 
As $\tau$ increases well beyond the range of Fig.~\ref{fig:g0inv}, 
$\xi_\infty$ approaches a molecular size and the anisotropic stress becomes ill-defined. 
Then,  because of the background contribution to the
osmotic susceptibility \cite{seng,swinney}, 
the partial derivative in Eq.~(\ref{eqn:mob}) should stop increasing in proportion to $\tau^\gamma$.
The mobility should  then be explained in terms of
the conventional mechanism rather than the PA.
\par

It is demonstrated in Ref.~\cite{mood} that, for nanocolloids in a nonelectrolyte solution, the diffusiophoretic mobility in the conventional theory
can also be obtained by molecular dynamics simulation.
We naively convert the second result from the right end 
in their Fig.~3(c) to ${\cal M}= 1.6\times 10^2\ \mu$m$^2/$s 
for a mixture of LW by taking 2,6-lutidine  as the component a and by 
using $k_{\rm B}T_{\rm cr}$ and $\xi_0$ as the unit energy and length, respectively. 
The potential for the interaction between the nanocolloid and the solute
is mentioned in Ref.~\cite{mood}.  Regarding it as representing the adsorption of the component a
and assuming the length of the interaction range to be the unit length,
we tentatively convert its value at the surface of the immobile region to
$h= 1.0\ $cm$^3/$s$^2$.  {Noting that the converted values of ${\cal M}$ and $h$
are linked via the unit energy and} neglecting the {size} dependence of the mobility, 
we can deduce that the values of ${\cal M}$ shown by the cross and square in Fig.~\ref{fig:g0inv}(b) of the present study
would approximately become $60$ and $10\ \mu$m$^2/$s, respectively, in the range of $\tau$
validating the conventional theory. 
It is thus possible, for $h>0$, that the mobility for large $h$ is made
smaller inside the critical regime than outside 
by larger osmotic susceptibility, and that the mobility for small $h$
is made larger inside the critical regime by thicker adsorption layer.
This possibility clearly needs further investigation, particularly because
the converted values are highly dependent on the unit length.  
\par

In this Letter, we theoretically show 
for diffusiophoresis in a nonelectrolyte fluid mixture near the demixing critical point that 
the mobility can have
a large enough magnitude to be detected {only in the critical regime}, that   
its dependence on temperature can be changed qualitatively by
the surface field, and that its dependence on particle radius can remain explicit
even when the radius exceeds $0.1\ \mu$m.    \\


\noindent
{\bf Acknowledgements}\\
The author thanks Dr.~S. Yabunaka for stimulating discussion.\\


\noindent
{\bf Supplemental Material}\\
{Here}, we use the notation {of} the main text, {and 
most of the contents in the first three paragraphs are described in more detail   
in Ref.~\cite{yabufuji}.}  In Ref.~\cite{OkamotoOnuki}, the coarse-grained FEF
is derived from the Ginzburg-Landau-Wilson type of bare model and
contains $\xi_0$, $C_1$ and $C_2$ as material constants.  
{We obtain $C_2/(C_1\xi_0)=2\pi^{2}/3$ from the 
renormalization-group calculation at the one loop order.}
{We define $\omega$ as} $(\xi_0/\xi)^{1/\nu}$, {where $\xi$ is the local correlation length,}
 and ${\hat\omega}$ as $\omega/\tau_*{\equiv (r_0/\xi)^{1/\nu}}$, {where
$r_0$ is the particle radius.}
A dimensionless order parameter {${\hat \psi}({\hat r})$}
is defined as
$\psi({r} )/\psi_*$, where $\psi_*$ denotes 
$\tau_*^\beta/\sqrt{C_2}$
{and the dimensionless radial distance ${\hat r}$ equals $r/r_0$.} 
Introducing $\mu_*\equiv 
{k_{\rm B}{T}C_1\xi_0/(C_2r_0^3 \psi_*)}$, 
we {define} dimensionless {chemical potential and surface field as} 
${\hat \mu}\equiv \mu/\mu_*$ and ${\hat h}\equiv h /\left(\mu_*r_0\right)$,
{respectively.  We have}
 $K(\psi){=}k_{\rm B}T C_{1}\omega^{-\eta\nu}$, {and}
$f_{\rm b}(\psi){=}\mu_*\psi_*{\hat f}_{\rm b}(\psi/\psi_*)$ with
${\hat f}_{\rm b}$ being defined as
 \begin{equation}
{\hat f}_{\rm b}({\hat\psi})={\hat\omega}^{\gamma-1}{\hat \tau}
\left( \frac{{\hat\psi}^2}{2}+\frac{{\hat\psi}^4}{12{\hat\omega}^{2\beta-1}{\hat \tau}}\right)
\ .\label{eqn:prefmm}\end{equation}
{Hereafter,} ${\hat\omega}$ is 
regarded as a function of ${\hat \psi}$
because {of} a self-consistent condition,
${\hat \omega}={\hat \tau}+{\hat \omega}^{1-2\beta}{\hat \psi}^{2}$.
{We have $\omega=\tau$ for $\psi=0$, and have
$\xi=r_0$ for $\psi=\psi_*$ at $\tau=0$.}
\par

The {order parameter in the reference state} $\psi^{(0)}$, vanishing far from the particle, 
equals $\psi$ minimizing ${\cal F}[\psi]$ under $\mu=0$.  The stationary condition yields
\begin{equation}
{\hat f}_{\rm b}'
=-\frac{\eta\nu}{2}\frac{{\hat \omega}'}{{\hat\omega}^{\eta\nu+1}}\left(\partial_{\hat r}{\hat\psi}\right)^2
+\frac{1}{{\hat \omega}^{\eta\nu}}\left(\partial_{\hat r}^2+
\frac{2}{{\hat r}}\partial_{\hat r}\right){\hat\psi} \label{eqn:profile}
\end{equation} for ${\hat r}>1$, together with  
$\partial_{\hat r}{\hat\psi}=- {\hat h} {\hat \omega}^{\eta\nu}$ at ${\hat r}=1$.
Thus, ${\hat \psi}^{(0)}$ is totally determined by ${\hat \tau}$ and ${\hat h}$. 
{The definition of} $\Phi$ {leads to}
$\Phi({\hat r})= - {\hat r}^2  \partial_{\hat r} {\hat \psi}^{(0)}({\hat r}) /{\sqrt{5\pi}}$;
${\cal A}({\hat r})$ and ${\cal B}({\hat r})$ 
are {respectively given by}
$L_*/L(\psi^{(0)}(r_0{\hat r}))={\hat f}_{\rm b}''/{\hat \omega}^\nu$
and
\begin{equation}
\sqrt{5\eta_{\rm s}L_*}\frac{L'(\psi^{(0)}(r_0{\hat r}))}{r_0 L(\psi^{(0)}(r_0{\hat r}))}
=\sqrt{{\frac{5\pi}{9}}}\left(\nu \frac{{\hat\omega}'}{{\hat\omega}}-\frac{{\hat f}_{\rm b}'''}{{\hat f}_{\rm b}''}\right)   \ . 
\label{eqn:Bdef}\end{equation}
{No diffusive flux at the particle surface gives} ${\cal Q}'{(1)}=0$.
We obtain $v_\theta^{(1)}/U=-\sin{\theta}[\partial_{\hat r} {\hat r}^2{\cal R}({\hat r})]/(2{\hat r})$ {using}
the incompressibility condition.   {The no-slip boundary condition at the surface gives}
${\cal R}{(1)}=1$ and ${\cal R}'{(1)}=0$.
Far from the particle, ${\cal R}$ vanishes and ${\cal Q}({\hat r})$ becomes $g{\hat r}$, {which is consistent with
Eqs.~(\ref{eqn:calQ}) and (\ref{eqn:calR}) because $\Phi$ vanishes there.
Notably}, $q_0$ satisfies the boundary conditions for ${\cal Q}$ and deletes the left-hand side (LHS) of
{Eq.~(\ref{eqn:calQ}).}  
\par

{In the main text, we modify Eqs.~(\ref{eqn:calQ}) and (\ref{eqn:calR}) by introducing $\kappa$.
Defining ${\cal I}$ and ${\cal J}$
so that the modified RHSs respectively equal   
$\kappa {\cal I}({\hat r})$ and $\kappa {\cal J}({\hat r})$,  we obtain
\begin{equation}
{\tilde {\cal Q}}_\kappa\left({\hat r} \right)=q_0({\hat r})-\frac{\kappa}{3}\int_1^\infty d{\hat s}\ 
\varGamma_{{\rm Q}}\left({\hat r}, {\hat s}\right){\cal I}({\hat s})\ ,
\label{eqn:inteq-Q}
\end{equation}
where $\varGamma_{\rm Q}(r, s)$ 
is given by $(r^{-2}s^{-2}/2)+r^{-2}s$ for $r\ge s$ and equals $\varGamma_{\rm Q}(s, r)$, and  
\begin{equation}
{\tilde {\cal R}}_\kappa \left({\hat r}\right)=
1+\alpha_0({\hat r})
-\frac{\kappa}{30}\int_1^\infty d{\hat s}\ \varGamma_{{\rm R}}\left({\hat r}, {\hat s}\right){\cal J}({\hat s})
\ .\label{eqn:inteq-R}\end{equation}
Here, $\varGamma_{\rm R}(r, s)$ 
is given by the sum of $r^{-3}s^2-5r^{-1}$ and
$(15r^2s^2-5r^2-5s^2+3)/(2r^3s^3)$
for $r\ge s$, and equals $\varGamma_{\rm R}\left(s,r\right)$.
Equations (\ref{eqn:inteq-Q}) and (\ref{eqn:inteq-R}) are the integral equations mentioned above Eq.~(\ref{eqn:qn}).  
With ${\mathsf E}$ denoting the rate-of-strain tensor, the drag force is given by the  
area integral of the dot product of $-p{\mathsf 1}-{\mathsf \Pi}+2\eta_{\rm s} {\mathsf E}$ and ${\v e}_r$ over the particle surface;
$f_{\rm s}$ does not contribute to the drag coefficient $\gamma_{\rm d}$ \cite{effvis}.
The part of the integral involving ${\mathsf \Pi}$ is rewritten by using
Gauss' divergence theorem and
 $\nabla\cdot{\mathsf \Pi}=\varphi\nabla\mu$, whereas the other part by using
Eq.~(\ref{eqn:inteq-R}) for $\kappa=1$.  As a result, we obtain
the expression of the deviation ratio $\Delta{\hat \gamma}_{\rm d}$ mentioned in the main text.}
\par

{Because} the differences of the solutions {${\cal Q}$ and ${\cal R}$} 
for $g=0$ subtracted from the respective solutions for a value of $g$ 
are proportional to $g$, {$\Delta{\hat \gamma}_{\rm d}$} is a linear function of $g$.  
{Because} 
${\cal Q}$ and ${\cal R}$
are totally determined by $g$, ${\hat \tau}$
and ${\hat h}$, {$g_0$} is totally determined by ${\hat \tau}$
and ${\hat h}$.
If the sign of $h$ is changed, the signs of
$\Phi$ and ${\cal B}$ are changed, whereas ${\cal A}$ is unchanged.
Then, if that of $g$ is also changed, that of ${\cal Q}$ is changed, whereas ${\cal R}$ and $\Delta{\hat \gamma}_{\rm d}$
are unchanged.  Thus, $g_0$ is an odd function of $h$.  
\par

For a mixture of LW, $C_2 = 7.14 \times 10^{-7}\ $m$^6/$kg$^2$ is obtained in Ref.~\cite{pipe}  from the experimental data
\cite{jaya, mirz, to} with the aid of Eqs.~(3.3) and (3.23) of Ref.~\cite{OkamotoOnuki}.
This value yields $\psi_*=4.70\times 10\ $kg$/$m$^3$ for $r_0=0.1\ \mu$m.
{Onto the surface of} a silica particle, 2,6-lutidine 
preferentially adsorbs \cite{Omari}. {In this case, regarding}  $\eta_{\rm s}$ as a constant 
gives a good approximation 
in the range of $\tau$ examined, thanks to the background contribution \cite{pipe}.
We use $\eta_{\rm s}=2.43\ $mPa$\cdot$s from the data in Refs.~\cite{gratt,stein}.
\par

\begin{figure}
\includegraphics[width=7cm]{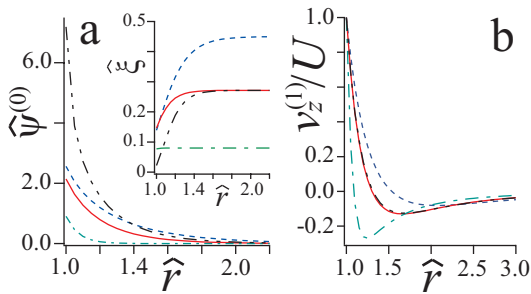}
\caption{Against the dimensionless radial distance ${\hat r}$, {we plot the dimensionless order parameter in the reference state}
${\hat \psi}^{(0)}({\hat r})$ in (a), {the dimensionless correlation length} ${\hat \xi}$ in the reference state in the inset, and
{the $z$ component of the velocity field divided by that of the particle velocity} $v_z^{(1)}/U$ at $\theta=\pi/2$ in (b) 
{by setting the particle radius} $r_0$ {equal to} $0.1\ \mu$m. 
The blue dashed, red solid, and green dash-dot curves are obtained {by setting the reduced temperature} $\tau$ {equal to}
$1.82\times 10^{-4}, 4.05\times 10^{-4}$, and 
$2.82\times 10^{-3}$, respectively, under $h=1.55\times 10^{-2}\ $cm$^3/$s$^2$.
The black dash-dot-dot curve is obtained for $\tau= 4.05\times 10^{-4}$ and $h=1.73\times 10^{-1}\ $cm$^3/$s$^2$. }
\label{fig:psivelz}
\end{figure}

We obtain ${\hat \psi}^{(0)}$ by numerically solving Eq.~(\ref{eqn:profile}) and the boundary conditions mentioned around this equation.
In Fig.~\ref{fig:psivelz}(a), ${\hat \psi}^{(0)}{({\hat r})}$ decreases as ${\hat r}$ increases.  As
$\tau$ decreases, the decrease is more gradual, {{\it i.e.\/},} the adsorption layer becomes thicker. 
The value of ${\hat \psi}^{(0)}(1)$ increases {with} $h$.
We define ${\hat \xi}$ as $\xi/r_0\equiv {\hat \omega}^{-\nu}$,
calculate its values in the reference state with the aid of the self-consistent condition, and plot the results in the inset of Fig.~\ref{fig:psivelz}(a). 
{The PA causes ${\hat \xi}<\xi_\infty/r_0={\hat \tau}^{-\nu}$} near the particle surface.
For the force-free particle motion ($g=g_0$),
we plot  $v_z^{(1)}/U=-v_\theta^{(1)}/U$ at $\theta=\pi/2$
in Fig.~\ref{fig:psivelz}(b).  
For a set of $(\tau, h)$, the plotted function changes {its sign} from positive to negative
as ${\hat r}$ increases.  Its zero 
is larger as $\tau$ decreases
for $h=1.55\times 10^{-2}\ $cm$^3/$s$^2$.  {At} $\tau=4.05\times 10^{-4}$, 
the curves for the different values of $h$ approximately coincide with each other.
{This property} is also observed for each of the other values of $\tau$
although data not shown, {and would appear because} $\xi_\infty$ indicates the thickness of the adsorption layer. 
{As} required in our coarse-grained description, each curve in Fig.~\ref{fig:psivelz}(b) can be well traced even if viewed
with the local resolution provided by the corresponding
local value of ${\hat \xi}$ shown in the inset of Fig.~\ref{fig:psivelz}(a). 
\par

We write ${\bar v}_{n}$ for the partial volume per unit mass of the component $n$, and
write ${\bar v}_\pm$ for $({\bar v}_{\rm a}\pm{\bar v}_{\rm b})/2$.
If the mass densities are homogeneous, we have
\begin{equation}
\left.\frac{\partial\mu}{\partial \varphi}\right)_{T\rho}= \left.\frac{\partial\mu}{\partial P}\right)_{Tc_{\rm a}}
 \left.\frac{\partial P}{\partial \varphi}\right)_{T\rho} 
+\left.\frac{\partial\mu}{\partial c_{\rm a}}\right)_{TP}
 \left.\frac{\partial c_{\rm a}}{\partial \varphi}\right)_{T\rho}
\ .\end{equation}
The LHS above equals $f_{\rm b}''(\psi)$, whereas
the first and second derivatives of the first term on the RHS 
equal ${\bar v}_-$ and $f_{\rm b}''(\psi)\varphi$, respectively. 
Thus, with the aid of the identity $1={\bar v}_{\rm a}\rho_{\rm a} +{\bar v}_{\rm b}\rho_{\rm b}$, we obtain
 \begin{equation}
 \left.\frac{\partial\mu}{\partial c_{\rm a}}\right)_{TP} =2\rho_{\rm c}^2{\bar v}_{+{\rm c}} f_{\rm b}''(0)
\end{equation}
far from the particle in the reference state. 
This can be substituted into {Eq.~(\ref{eqn:mob}).}
From the data in Table III of Ref.~\cite{jaya}, we estimate 
${\bar v}_{\rm a}=1.03\times 10^{-3}\ $m$^3/$kg and  ${\bar v}_{\rm b}=1.00\times 10^{-3}\ $m$^3/$kg for a mixture of LW
lying in the homogeneous phase near the demixing critical point \cite{pipe}.  Thus, the mixture has ${\bar v}_{+{\rm c}}\approx 1/\rho_{\rm c}$.
Because of $f_{\rm b}''(0)=
k_{\rm B}TC_1\tau^\gamma/\xi_0^2$, we have $L_*=(\xi_0/r_0)^\eta r_0/(6\pi\eta_{\rm s}C_1)$.
\par

{For data points showing ${\cal M}\approx 50\ \mu$m$^2$/s in Fig.~\ref{fig:g0inv},
$h\psi^{(0)}(r_0)$ should be larger than $6.63\times 10^{-7}\ $kg/s$^2$, which value
is calculated from ${\hat \psi}^{(0)}(1)$ for $\tau=2.82\times 10^{-3}$
and $h=1.55\times 10^{-2}\ $cm$^3$/s$^2$ in Fig.~\ref{fig:psivelz}(a). 
Thus, the condition $\eta_{\rm s} \left|\varepsilon U \right|\ll \left|h\psi^{(0)}(r_0)\right|$, mentioned in the
main text, is satisfied if the particle speed is much smaller than $3\times 10^2\ \mu$m/s 
for the data points mentioned above.
Considering that ${\hat\psi}^{(0)}(1)$ can be roughly approximated  to be unity in Fig.~\ref{fig:psivelz}(a),
we can nondimensionalize the particle speed by dividing it 
by $h\psi_*/\eta_{\rm s}$ and use the dimensionless quotient
 as the smallness parameter.}


\begin{thebibliography}{99}
\bibitem{derja} B. V. Derjaguin, S. S. Dukhin, and V. V. Koptelova,  
``Capillary osmosis through porous partitions and properties of
boundary layers of solutions," 
J. Coll.~Interf.~Sci.~{\bf 38}, 584--595 (1972).
\bibitem{andersJFM}  J. L. Anderson, M. E. Lowell, and D. C. Prieve, 
``Motion of a particle generated by chemical
gradients part 1. non-electrolytes,'' J. Fluid Mech.~{\bf 117}, 107--121 (1982).
\bibitem{SepPurMeth}   J. L. Anderson and  D. C. Prieve,
``Diffusiophoresis: migration of colloidal particles in
gradients of solute concentration,'' Sep.~Purif.~Methods {\bf 13}, 67--103 (1984).
\bibitem{prieve} D. C. Prieve and R. Roman, ``Diffusiophoresis of a rigid sphere through a viscous
electrolyte solution,'' J. Chem.~Soc., Faraday Trans.~2, {\bf 83}, 1287--1306 (1987).
\bibitem{anders} J. L. Anderson,
``Colloid tranport by interfacial forces," 
Ann.~Rev.~Fluid Mech.~{\bf 21},  61--99 (1989).
\bibitem{staff} P. O. Staffeld and J. A. Quinn, ``Diffusion-induced banding of colloid particles via diffusiophoresis
2. non-electrolyte,'' J. Coll.~Interf.~Sci.~{\bf 130}, 88--100 (1989).
\bibitem{abecas} B. Ab{\'e}cassis, C. Cottin-Bizonne, C. Ybert, A. Adjari and L. Bocquet, ``Osmotic manipulation of particles for microfluidic
applications,'' New J. Phys.~{\bf 11}, 075022 (2009).
\bibitem{volpe}{G. Volpe, I. Buttinoni, D. Vogt, H.-J. K{\"  u}mmerer, and C. Bechinger, ``Microswimmers in patterned environments,''
Soft Matter {\bf 7}, 8810 (2011).}
\bibitem{osmflow} C. Lee, C. Cottin-Bizonne, A.-L. Biance, P. Jpseph, L. Bocquet, and C. Ybert,
``Osmotic flow through fully permeable nanochannels,"  
Phys.~Rev.~Lett.~{\bf 112}, 244501 (2014). 
\bibitem{kar}  A. Kar, T.-Y. Chang, I. O. Riviera, A. Sen, and D. Velegol, ``Enhanced transport into and out
of dead-end pores,'' ACS Nano {\bf 9}, 746--753 (2015).
\bibitem{shinPNAS} S. Shin, E. Urn, B. Sabass, J. T. Auit, M. Rahimi, P. B. Warren, and H. A. Stone, 
``Size-dependent control of colloid transport via solute
gradients in dead-end channels,''
Proc.~Natl.~Acad.~Sci.~USA {\bf 113}, 257--261 (2016).
\bibitem{keh} H. J. Keh,  
``Diffusiophoresis of charged particles and diffusioosmosis of electrolyte solutions,"  
Curr.~Opin.~Coll.~Interf.~Sci.~{\bf 24}, 13--22 (2016).
\bibitem{veleg} D. Velegol, A. Garg, R. Gusha, A. Kar, and M. Kumar, ``Origins of concentration gradients for
diffusiophoresis,'' Soft Matter {\bf 12}, 4686--4703 (2016).
\bibitem{marb} S. Marbach and L. Bocquet, 
``Osmosis, from molecular insights to large-scale applications," 
Chem.~Soc.~Rev.~{\bf 48},  3102--3144 (2019).
\bibitem{shin} S. Shin, 
``Diffusiophoretic separation of colloids in microfluidic flows," 
Phys. Fluids {\bf 32}, 101302 (2020).
\bibitem{frenkel} {S. Ram{\' i}rez-Hinestrosa and D. Frenkel, ``Challenges in modelling diffusiophoretic transport,''
Eur.~Phys.~J. B {\bf 94}, 199 (2021).}
\bibitem{vinze}
{P. M. Vinze, A. Choudhary, and S. Pushpavanam, ``Motion of an active particle in a linear
concentration gradient,''  Phys.~Fluids {\bf 33}, 032011 (2021).}
\bibitem{okafujiko}
R. Okamoto, Y. Fujitani, and S. Komura, 
``Drag coefficient of a rigid spherical particle in a near-critical binary fluid mixture,"  
J. Phys.~Soc. Jpn  {\bf 82},  084003  (2013).
\bibitem{furu}  A. Furukawa, A. Gambassi, S. Dietrich, and H. Tanaka, 
``Nonequilibrium critical Casimir effect in binary fluids,"
Phys.~Rev.~Lett.~{\bf 111}, 055701 (2013).
\bibitem{Cahn} J. W. Cahn, 
``Critical point wetting,"
  J.~Chem.~Phys.~{\bf 66},  3667--3672 (1977).
\bibitem{beyslieb}
D. Beysens and S. Leibler, 
``Observation of an anomalous adsorption in a critical binary mixture," 
J. Physique Lett.~{\bf 43},  133--136 (1982).
\bibitem{binder}
M. N. Binder, in {\it Phase Transitions and Critical
  Phenomena VIII\/}, edited by C.~Domb and J.~Lebowitz,
``Critical behavior at surfaces''  
(Academic, London, 1983).
\bibitem{beysest}
D. Beysens, and D. Est{\`e}ve, 
``Adsorption phenomena at the surface of silica spheres in a binary liquid mixture,"
  Phys.~Rev.~Lett.~{\bf 54},  2123--2126  (1985).
\bibitem{holyst} R. Holyst and A. Poniewierski,  
``Wetting on a  spherical surface,''  
Phys.~Rev.~B  {\bf 36},  5628--5630 (1987).
\bibitem{diehl97}
H. W. Diehl, 
``The theory of boundary critical phenomena,"   
Int.~J. Mod.~Phys.~B  {\bf 11}, 3503--3523 (1997).
\bibitem{beys2019}  D. Beysens,
``Brownian motion in strongly fluctuating liquid,''  
Thermodyn.~Interf.~Fluid Mech.~{\bf 3},  1--8 (2019).
\bibitem{fujitani2021} Y. Fujitani, 
``Suppression of viscosity enhancement around a Brownian particle in a near-critical binary fluid mixture,''
J. Fluid Mech. {\bf 907}, A21 (2021).
\bibitem{yabufuji} S. Yabunaka and Y. Fujitani,
``Drag coefficient of a rigid spherical particle in a near-critical binary fluid mixture, beyond the regime of the Gaussian model,"
J. Fluid Mech.~{\bf 886} A2 (2020).
\bibitem{fisher-auyang} M. E. Fisher, and H. Au-Yang, 
``Critical wall perturbations and a local free energy functional,"
Physica A~{\bf 101}, 255--264 (1980).
\bibitem{OkamotoOnuki} R. Okamoto and A. Onuki, 
``Casimir amplitude and capillary condensation of near-critical binary fluids between parallel plates: renormalized local functional theory,"
  J. Chem.~Phys.~{\bf 136}, 114704 (2012).
\bibitem{Yabu-On}
S. Yabunaka and A. Onuki, 
``Critical adsorption profiles around a sphere and a cylinder in a fluid at criticality: local functional theory,"  
Phys.~Rev.~E  {\bf 96},  032127  (2017).
\bibitem{yabuokaon} S. Yabunaka, R. Okamoto, and A. Onuki, 
``Hydrodynamics in bridging and aggregation of two colloidal particles in a near-critical binary mixture," 
Soft Matter {\bf 11}, 5738--5747 (2015).
\bibitem{undul} Y. Fujitani,
``Undulation amplitude of a fluid membrane
 in a near-critical binary fluid mixture calculated beyond the Gaussian model
supposing weak preferential attraction,''
J. Phys.~Soc.~Jpn.~{\bf 86},  044602 (2017).
\bibitem{peli}
A. Pelisetto and E. Vicari, 
``Critical phenomena and renormalization-group theory,"  
Phys.~Rep.~{\bf 368},  549--727 (2002).
\bibitem{wolynes} P. G. Wolynes, "Osmotic effects near the critical point," J. Phys. Chem. {\bf 80},  1570--1572 (1976).
\bibitem{pipe}
S. Yabunaka and Y. Fujitani, 
``Isothermal transport of a near-critical binary fluid mixture through a capillary tube with the preferential adsorption,'' 
Phys.~Fluids {\bf 34}, 052012 (2022).
\bibitem{Hohen-Halp}
P. C. Hohenberg and B. I. Halperin,
``Theory of dynamic critical phenomena,''  
Rev. Mod. Phys.~{\bf 49}, 435--479 (1977).
\bibitem{onukibook}  A. Onuki, {\it Phase Transition Dynamics\/}
  (Cambridge University Press, Cambridge, 2002), Chap.~6.1.
\bibitem{onukiPRE} A. Onuki, ``Dynamic equations and bulk viscosity near the gas-liquid critical point,'' Phys.~Rev.~E {\bf 55}, 403--420 (1997).
\bibitem{kawasaki}  K. Kawasaki, ``Kinetic equations and time correlation functions of critical fluctuations,'' Ann.~Phys.~({\it N. Y.}) {\bf 61}, 1--56 (1970).
\bibitem{seng} J. V. Sengers, 
``Transport properties near critical points,''
Int.~J. Thermophys.~{\bf 6}, 203--232 (1985).
\bibitem{stokes} G. G. Stokes, ``On the effect of the internal friction of fluid on the the motion of pendulums,'' 
Trans.~Camb.~Phil.~Soc.~{\bf 9}, 8--93 (1851).
\bibitem{suther} W. Sutherland, ``A dynamical theory of diffusion for nonelectrolytes
and the molecular mass of albumin,'' Phil.~Mag.~{\bf 9}, 781--785 (1905).
\bibitem{eins} A. Einstein, ``\"{U}ber die von der
  molekularkinetischen theorie der w\"{a}rme
 geforderte bewegung von in
  ruhenden fl\"{u}ssigkeiten suspendierten teilchen,'' 
Ann.~Phys.~({\it Leipzig\/}) {\bf 322}, 549--560 (1905).
\bibitem{fpd} Y. Fujitani,
``Connection of fields across the interface in the fluid particle dynamics method for colloidal dispersion,''
 J.~Phys.~Soc.~Jpn.~{\bf 76}, 064401 (2007).
\bibitem{mirz} S. Z. Mirzaev, R. Behrends, T. Heimburg, J. Haller, and U. Kaatze, 
``Critical behavior of 2,6-dimethylpyridine-water: Measurements of specific heat, dynamic light scattering, and shear viscosity," 
J. Chem. Phys. {\bf 124} 144517 (2006).
\bibitem{swinney}  H. L Swinney and D. L. Henry, 
``Dynamics of fluids near the critical point: decay rate of order-parameter fluctuations,'' Phys.~Rev.~A  {\bf 8}, 2586--2617 (1973). 
\bibitem{mood} N. Sharifi-Mood, J. Koplik, and C. Maldarelli, ``Molecular dynamics simulation of the motion of colloidal nanoparticles in a
solute concentration gradient and a comparison to the continuum limit,'' Phys. Rev.~Lett.~{\bf 111}, 184501 (2013).
{\bibitem{effvis} Y. Fujitani, ``Effective viscosity of a near-critical binary fluid mixture with colloidal particles dispersed dilutely under weak shear,''
J. Phys.~Soc.~Jpn. {\bf 83}, 084401.}
\bibitem{jaya} 
Y. Jayalakshmi,  J. S. van Duijneveldt, and  D. Beysens, 
``Behavior of density and refractive index in mixtures of 2,6-lutidine and water,''   
J. Chem.~Phys.~{\bf 100}, 604--609 (1994). 
\bibitem{to} K. To, 
``Coexistence curve exponent of a binary mixture with a high molecular weight polymer,''
Phys.~Rev.~E {\bf 63}, 026108 (2001).
\bibitem{Omari}
R. A. Omari, C. A. Grabowski \& A. Mukhopadhyay,
``Effect of surface curvature on critical adsorption,''
Phys.~Rev.~Lett.~{\bf 103},  225705 (2009).
\bibitem{stein} A. Stein, S. J. Davidson, J. C. Allegra, and G. F. Allen, 
``Tracer diffusion and shear viscosity for the system 2,6-lutidine-water near the lower critical point," 
J. Chem.~Phys. {\bf 56} 6164--6168 (1972).
\bibitem{gratt} C. A. Grattoni, R. A. Dawe, C. Y. Seah, and J. D. Gray, 
``Lower critical solution coexistence curve and physical properties (density, viscosity, surface tension, and interfacial tension) of
2,6-lutidine$+$water,"  
Chem. Eng. Data, {\bf 38}, 516--519 (1993).
\end{thebibliography}
\end{document}